\documentclass[final]{ias2}

\usepackage{graphicx} 
\usepackage{multirow}
\usepackage{array} 

\usepackage{hyperref} 
\usepackage{lineno}
\begin{document}

\markboth{High Energy Neutrinos from Space}{Thomas K. Gaisser}

\title{High Energy Neutrinos from Space}

\author[sin]{Thomas K. Gaisser} 
\email{gaisser@bartol.udel.edu}
\address[sin]{Bartol Research Institute and Department of Physics and Astronomy, 
University of Delaware, Newark, DE 19716, USA}

\begin{abstract}
This paper reviews the status of the search for high-energy neutrinos from astrophysical
sources.  Results from large neutrino telescopes in water (Antares, Baikal) and ice (IceCube)
are discussed as well as observations
from the surface with Auger and from high altitude with ANITA.  Comments on IceTop,
the surface component of IceCube are also included.
\end{abstract}

\keywords{Neutrinos, cosmic rays}

\pacs{Appropriate pacs here}
 
\maketitle


\section{Introduction}

A principal motivation for finding and studying high-energy neutrinos from
space is to understand better the sources of cosmic rays and how they accelerate
particles to high energy.  This review is organized around two connections
between cosmic rays and neutrinos.  In the first place, it is likely that 
neutrinos will be produced at some level 
in interactions of accelerated particles with gas or radiation fields
in or near the cosmic-ray sources.  Examples are hadronic interactions in the
material near an expanding supernova remnant or photo-pion production in
the radiation fields inside the jets of an accreting black hole.  The main
approach in these cases is to look for an excess of neutrinos from a particular direction
in the sky above the background of atmospheric neutrinos.  Potential sources
may be selected for study according to the likelihood that production
of neutrinos is expected.  Such targeted searches may increase the discovery potential
compared to a survey of the whole sky.

Production of neutrinos is also expected by interactions of cosmic rays
as they propagate through the Universe.  Locally, neutrinos are produced
as cosmic-rays interact with gas in the interstellar medium~\cite{Stecker}.  The expected
level (which is quite low) can be calculated directly from the observed gamma-ray flux
from the same source, which traces the gas in the disk of the galaxy.
A more interesting question is the level of neutrino production by
cosmic rays of ultra-high energy (UHECR) as they propagate through
the cosmic microwave background (CMB).  Protons with energies
above $5\times 10^{19}$~eV are above the threshold for production
of pions on CMB photons~\cite{Greisen,Zatsepin}.  Neutrinos would be
produced when the pions decay.  UHE cosmic-ray nuclei
also lose energy during propagation in the CMB by photo-disintegration~\cite{Stecker1},
but the level of neutrino production from subsequent decay of 
spallation neutrons is lower~\cite{Allard}.

\begin{figure}[t!]
\begin{center}
\includegraphics[width=0.7\columnwidth]{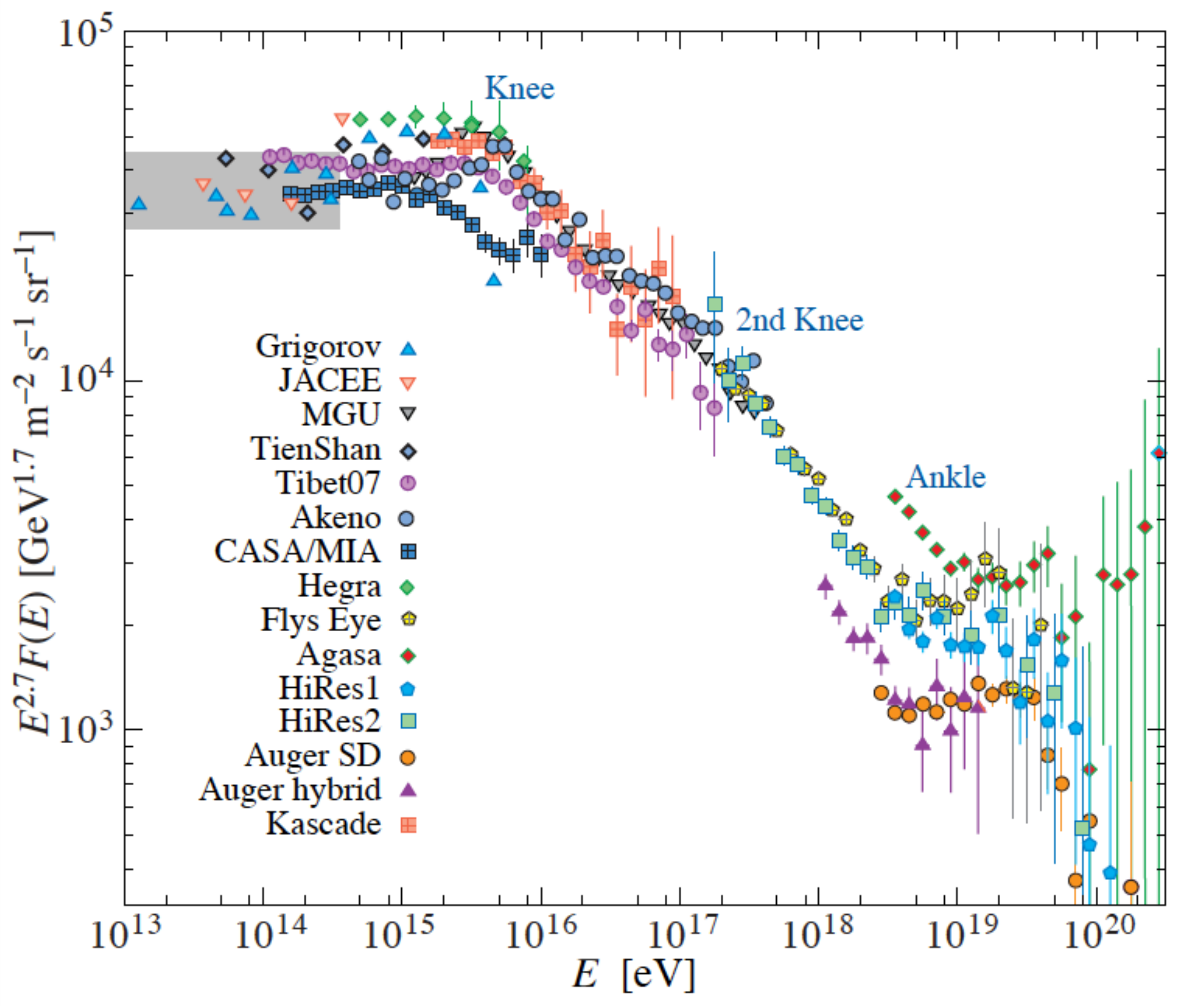}
\caption{Compilation of the primary cosmic-ray spectrum
measured by air shower experiments~\cite{RPP}.
}
\label{fig1}
\end{center}
\end{figure}

Figure~\ref{fig1} is a compilation of measurements of the
cosmic-ray spectrum at high energy.  The spectrum steepens between
$10^{15}$ and $10^{16}$~eV total energy per particle.  This feature, called
the knee, may reflect the decrease in the ability of galactic accelerators
such as supernova remnants to achieve such high energy.  The spectrum
flattens again around $3\times10^{18}$~eV~$=3$~EeV, a feature
known as the ankle.  It is generally assumed
that particles above this energy originate in more powerful
extragalactic sources.  Finally, the steepening of the spectrum
above $5\times10^{19}$~eV is usually assumed to be the effect of expected
energy losses during propagation through the CMB.  One reason to
search for cosmogenic neutrinos associated with propagation through
the CMB is to see if this assumption is correct.

\begin{table*}[thb]
\begin{center}
\begin{small}
\begin{tabular}{lcccl} \hline
Detector & Number of OMs & Enclosed volume  & Depth & Status \\ 
& & (Megatons) & (m.w.e.) & \\ \hline
Baikal~\cite{Baikal} (NT200+) & 230 & 10 & 1100-1310 & Operating \\
AMANDA~\cite{AMANDA} & 677 & 15 & 1350-1850 & 2000 - 2009 \\
ANTARES~\cite{Antares} & 900 & 10 & 2050-2400 & Operating \\
IceCube~\cite{IceCube} & 5160 + 324 & 900 & 1350-2280 & Operating \\
~~(IceCube DeepCore~\cite{DeepCore}) & (480) & (15) & (1950 - 2280) & Operating \\ \hline
KM3NeT~\cite{KM3NeT} & $\sim$10,000 & km$^3$ & 2300-3300  & Design study \\
&	&km$^3$& 3000-4000  & \\
& & km$^3$ & 1400-2400  & \\
GVD~\cite{GVD} (future Baikal) & $\sim$2500 & km$^3$ & 800-1300 & Design study \\ \hline
\end{tabular}
\end{small}
\caption{Parameters of existing and proposed neutrino telescopes 
ice.  Of the total of 5484 OMs in IceCube, 480 are deployed on DeepCore Strings and
324 in IceTop tanks.
The three depths listed for KM3NeT correspond to three possible locations,
NEMO~\cite{NEMO}, NESTOR~\cite{NESTOR} and Antares~\cite{Antares} in that order.}
\label{table1}
\end{center}
\end{table*}

\section{Detectors in ice and water}
The operating principle of detectors like IceCube, Antares and Baikal listed in
Table~\ref{table1} is the same as
Super-Kamiokande~\cite{SuperK} and SNO~\cite{SNO}:  events are reconstructed from times and amplitudes
in an array of optical sensors of Cherenkov light from charged
particles moving faster than the local speed of light in the detector.
The large detectors are, however, much less densely instrumented.
Compared to Super-K with 11,000 photomultiplier tubes (PMTs) in 40 kilotons
of water, IceCube has 5160 PMTs of half the diameter in a gigaton of ice.
The neutrino telescopes are optimized for large target volume with sparse instrumentation
to obtain the greatest sensitivity for relatively rare astrophysical neutrinos
of high energy.

\begin{figure}[thb]
\begin{center}
\includegraphics[width=0.9\columnwidth]{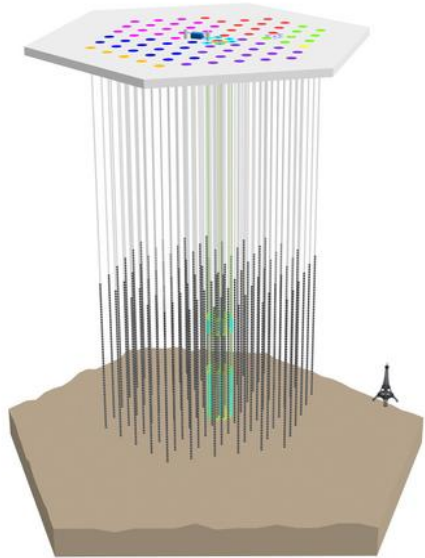}
\caption{Artist's drawing of the IceCube detector~\protect\cite{I3webpage}.}
\label{icecube}
\end{center}
\end{figure}

IceCube is currently the largest neutrino detector in operation~\cite{I3webpage}.  Construction
at the South Pole
was completed at the end of 2010, and the detector has been running since
May 20, 2011 with its full complement of 86 strings, each equipped with 60
digital optical modules (DOMs) at depths between 1450 and 2450 m in the ice.  Fig.~\ref{icecube} 
shows the completed detector, which also includes a surface air shower array
of 81 pairs of tanks, each instrumented with two DOMs and fully integrated into the
data acquisition system (DAQ) of IceCube.  The IceCube DOM 
includes, in addition to the 25~cm PMT~\cite{PMT}, 
a programmable data acquisition board~\cite{DAQ} that records
the amplitude as a function of time produced by photons hitting the photo cathode.
Digital signals are sent to the surface where computers build events from
physically related signals in the DOMs.  Times in individual DOMs are keyed to a single
GPS clock on the surface to an accuracy of $<3$~ns across the entire array including
IceTop.

Antares~\cite{Antares} is located in the Mediterranean Sea near Toulon.  It is the first
neutrino detector to operate in the open ocean, which requires deploying lines of
optical modules from a ship and continuously monitoring the positions of the
sensors as they respond to currents in the water.  Antares optical modules are
arranged in groups of three so that local coincidence can be used to overcome
the relatively high background of bioluminescence.  The Baikal detector also
operates in a natural body of water, but deployment occurs from the solid 
ice on the surface of the lake in winter.

\begin{figure}[thb]
\begin{center}
\includegraphics[width=0.5\columnwidth]{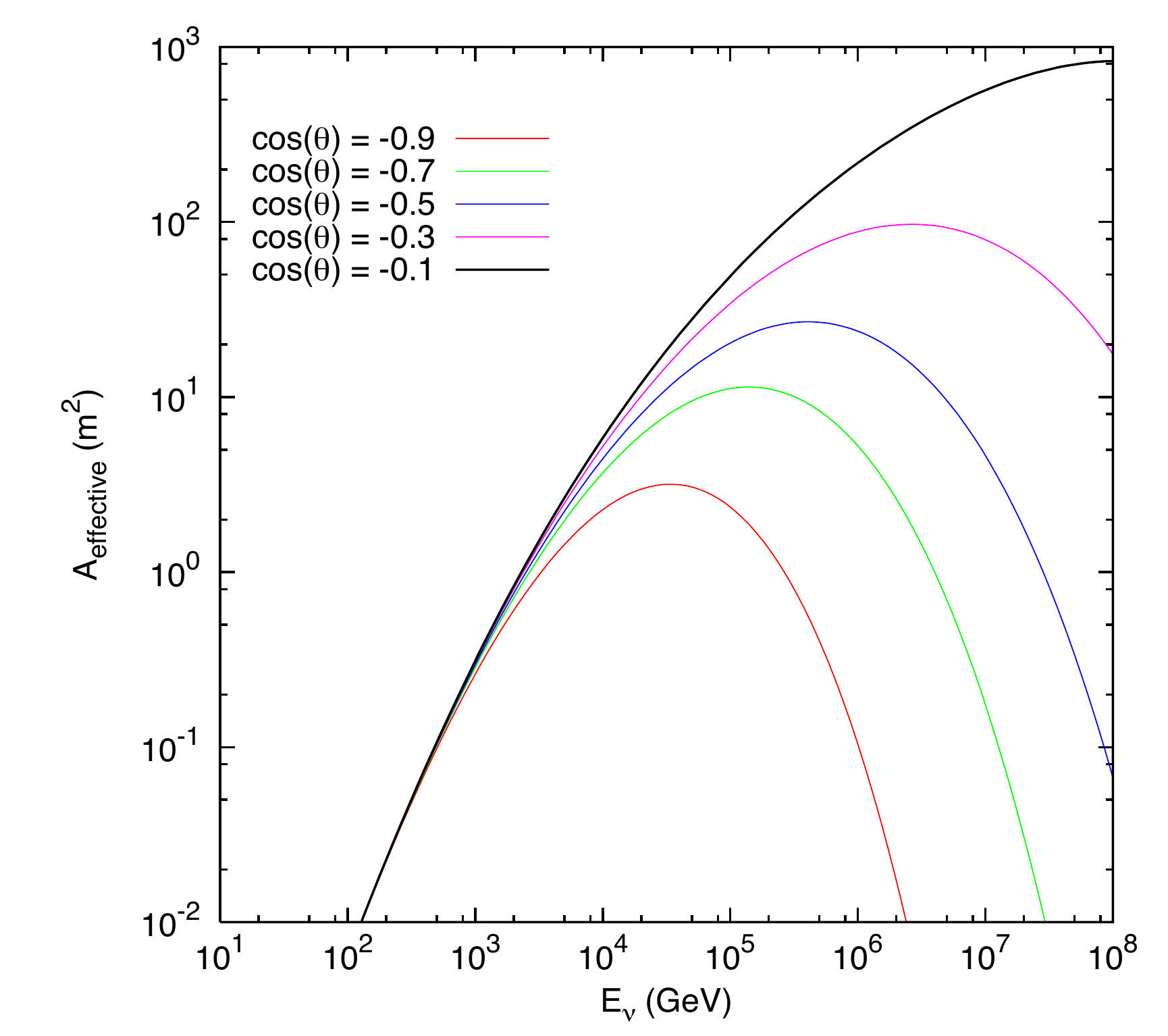}\includegraphics[width=0.5\columnwidth]{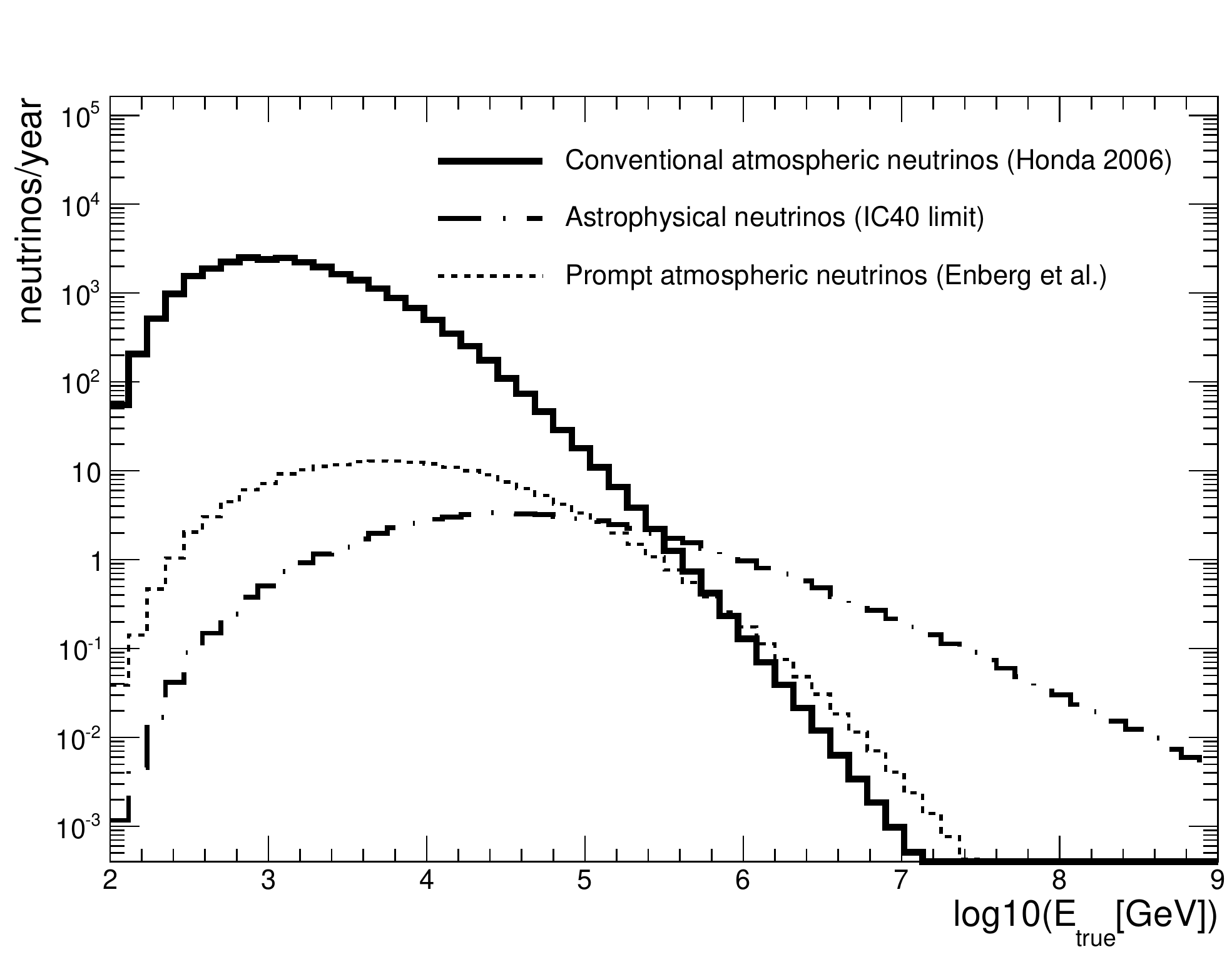}
\caption{Left: Effective area of an ideal cubic kilometer detector.
Right: Response of IceCube (with 59 strings in 2009-10) to three different
spectral shapes for muon neutrinos~\cite{Anne}. }
\label{Aeff}
\end{center}
\end{figure}

The muon channel is the most favorable in terms of
event rate in the TeV range and above because the target
volume is enlarged by the charged current interactions of
neutrinos outside the detector
that produce muons that go through the detector.
The most sensitive analysis uses the Earth as a shield
against the downward background of cosmic-ray muons
by selecting horizontal and
upward moving events.  For energies in the TeV range and
above, stochastic energy losses by muons become
important, and the light produced increases in proportion
to the muon energy according to the standard formula~\cite{RPP}
\begin{equation}
{{\rm d}E_\mu\over {\rm d}X}\,=\,-a -bE_\mu .
\label{Eloss}
\end{equation}
The event rate in this channel is a convolution of the neutrino
flux with the neutrino cross section, the detector response
and the range of the muon.  At high energy the Earth becomes
opaque to neutrinos, first for vertically upward neutrinos
($\sim 30$~TeV) and then for more horizontal events ($\sim$PeV).
Thus the effective area for $\nu_\mu$ in the charged current
channel is 
\begin{equation}
A_{\rm eff}(\theta,E_\nu)\,=\,\epsilon(\theta)\,A(\theta)\,P_\nu(E_\nu,E_{\mu})
\exp\{-\sigma_\nu(E_\nu)N_AX(\theta)\},
\label{Aeff2}
\end{equation}
where $X(\theta)$ is the slant depth (g/cm$^2$) along a zenith angle
$\theta > 90^\circ$, $N_A$ is Avogadro's number, $\sigma_\nu$ is the neutrino
cross section and $\epsilon(\theta)$ a reconstruction efficiency.
$$P_\nu(E_\nu,E_{\mu})\,=\,N_A\,\int_{E_\mu}^{E_\nu}\,{\rm d}E_\mu^*\,
{{\rm d}\sigma_\nu(E_\nu)\over {\rm d}E_\mu^*}\,R(E_\mu^*,E_{\mu})
$$
is the probability that a muon produced with energy $E_\mu^*$ reaches the detector with
energy $E_\mu$ sufficient to trigger the detector.  The muon range $R$
is calculated from Eq.~\ref{Eloss}.  The left panel of Fig.~\ref{Aeff} shows the
$\nu_\mu$ effective area for an ideal cubic kilometer detector with a threshold
$E_\mu=100$~GeV.  The neutrino rate is a convolution of effective area with neutrino flux.

In the TeV range and above muons typically pass through the detector,
so only a fraction of the muon energy contributes to light in the detector.
In this situation, simulations that incorporate the physics of neutrino 
interactions, of muon
energy loss and of ice properties must be used to relate the
measured light to the energy of the muon in the detector
and thence to the energy of the neutrino.
This is done either by convolving an assumed neutrino 
spectrum with the sequence $\nu_\mu\rightarrow\mu\rightarrow$~observed light,
or by an unfolding procedure.  An important feature of the analysis
is that the distribution of $\nu_\mu$ energies that give rise to
a given signal in the detector is different for the steep
atmospheric neutrino spectrum from what it would be for a
hard spectrum of astrophysical neutrinos.  This is illustrated in right panel of Fig.~\ref{Aeff},
which shows the distributions of neutrino energy that correspond respectively
to atmospheric neutrinos from decay of pions and kaons, to prompt 
atmospheric neutrinos from charm decay and to astrophysical neutrinos assumed
to have an $E^{-2}$ differential energy spectrum.  The responses are integrated
over $2\pi$ solid angle from below the horizon and use the full simulation of IC-59 for $A_{\rm eff}$.

\begin{figure}[thb]
\begin{center}
\includegraphics[width=0.8\columnwidth]{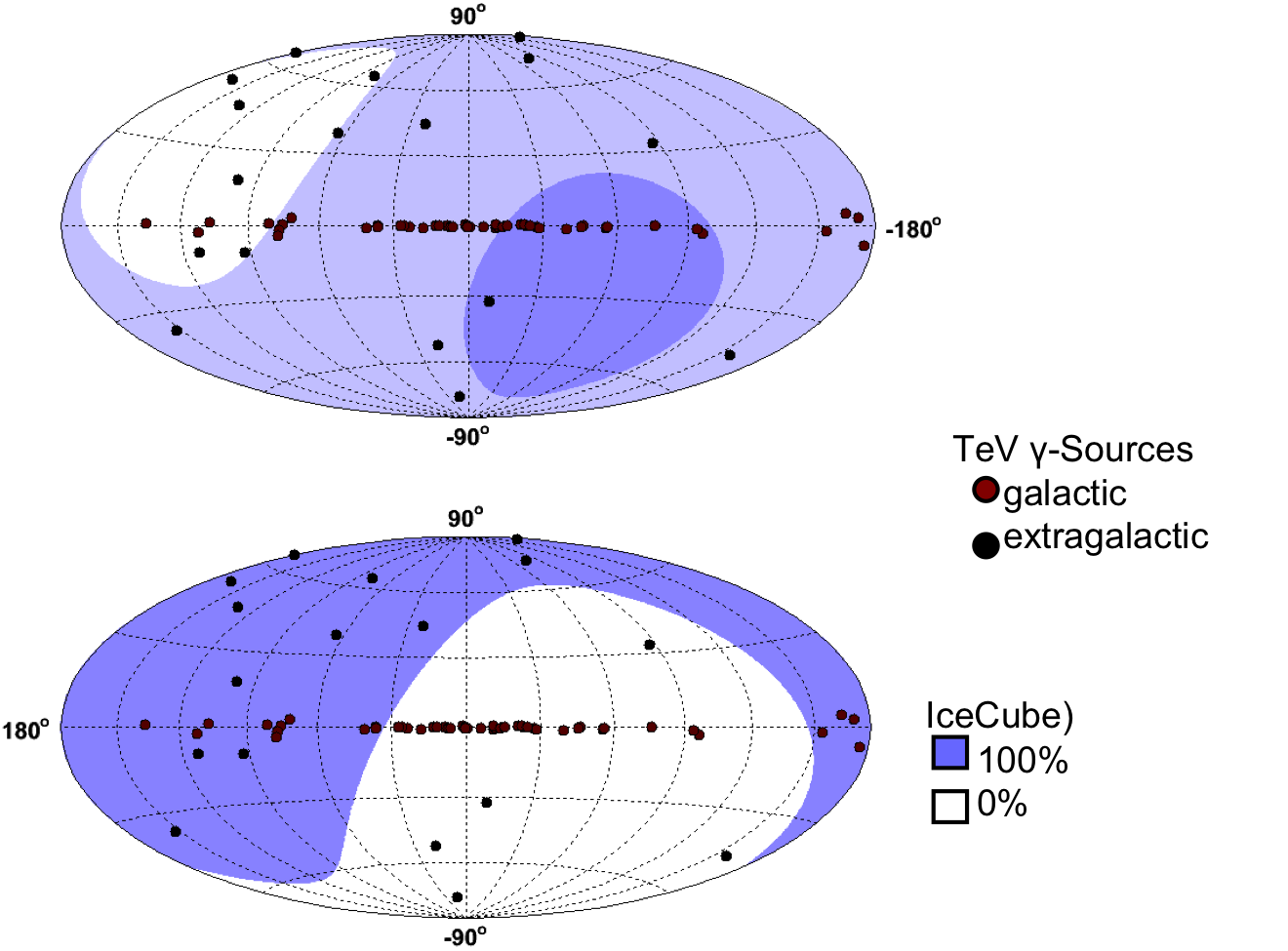}
\caption{Sky maps showing exposure of Antares (top) and IceCube (bottom)
for atmospheric $\nu_\mu$ from below the local horizon.  Figure
from Ref.~\cite{Gisela}.}
\label{skymaps}
\end{center}
\end{figure} 

\begin{figure}[bht]
\begin{center}
\includegraphics[width=0.65\columnwidth]{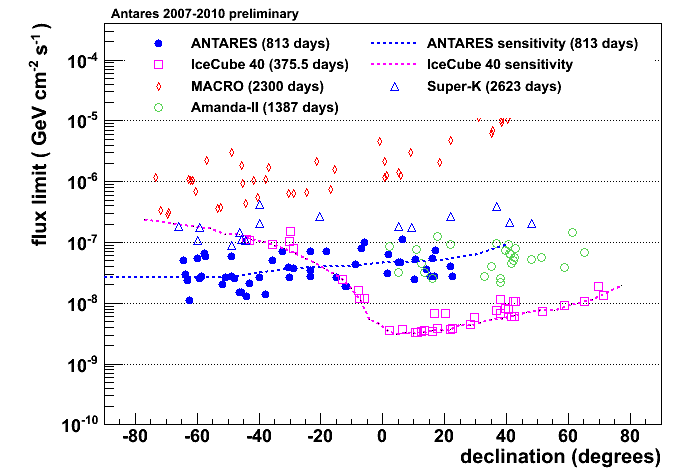}\includegraphics[width=0.35\columnwidth]{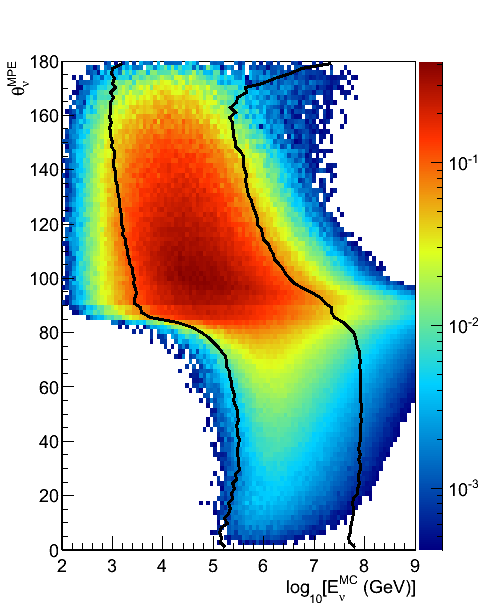}
\caption{Left: Sensitivity of Antares and IceCube (IC-40) for
point sources of $\nu_\mu$ (lines) and limits
for pre-selected sources (points) as a function of declination.
Figure from Ref.~\cite{AntaresPoint}.
Right: Color scale in angle and energy space for the neutrino point source search in IceCube
(not including DeepCore).}
\label{limits}
\end{center}
\end{figure} 

\section{Point source search}

The basic analysis in neutrino telescopes is to search for point sources of extraterrestrial
neutrinos using the direction of the neutrino-induced muon as a proxy for
the direction of the parent neutrino.  In the TeV range, the r.m.s. angle
between $\nu_\mu$ and $\mu$ is
$\sqrt{<\psi^2>}\,\approx\,1.8^\circ/\sqrt{E_\nu}(TeV)$.  The median angular
resolution for Antares is $0.5^\circ$~\cite{Gomez}.  For IceCube, 60\% of
the events are reconstructed to better than $1^\circ$~\cite{Teresa} based on analysis of
the moon shadow in cosmic-ray induced muons~\cite{Laura}.

An important difference between IceCube and the Antares location
at Northern mid-latitudes is exposure, as shown in Fig.~\ref{skymaps} from 
Ref.~\cite{Gisela}.  In particular, the central region of the
Galaxy is not visible in upward going events from the South Pole.  The
sensitivity of Antares and IceCube 
for point sources is shown by
the lines in Fig.~\ref{limits}.  In this plot the sensitivity is
shown for IceCube running with 40 strings installed in 2008-2009 (IC-40).
The individual points are limits
on preselected sources, which include galactic objects including supernova
remnants as well as potential extragalactic sources such as active galactic 
nuclei (AGN).  The point-source list for IC-40 includes 13 galactic
sources and 30 extragalactic sources.  Among the galactic sources targeted are 
several supernova remnants.  The extragalactic candidates are mostly active
galaxies.

For the Northern sky the larger IceCube detector
is by far the most sensitive.
Upper limits on specific point sources of neutrinos in the Northern sky
from IceCube are currently less than $10^{-11}$~cm$^{-2}$s$^{-1}$TeV$^{-1}$.
  The sensitivity
to point sources is approaching the level 
of $10^{-12}$~cm$^{-2}$s$^{-1}$TeV$^{-1}$~\cite{ptsrc} 
at which TeV gamma rays are seen from some blazars (e.g. Mrk 401~\cite{Daniel}).

IceCube extends the search for point sources to Southern declinations by
raising the energy threshold to reduce the high background of downward
atmospheric muons.  The sensitivity and limits from IC-40 from the Southern
sky are included in the left panel of Fig.~\ref{limits}.  The
right panel of Fig.~\ref{limits} shows the energy
response of IceCube with 59 strings (IC-59) as a function of declination.
For upward neutrinos from the Northern hemisphere the limits apply to 
$\nu_\mu$ energies in the TeV to PeV energy range.  For events from
above in IceCube (Southern sky) the relevant energy range
is approximately two orders of magnitude higher (100 TeV to 100 PeV).

With data from 2010 and later it will be possible to extend the search
for neutrinos in the Southern hemisphere to lower energy by using
the DeepCore subarray of IceCube.
In the final construction years of IceCube eight specially equipped strings
were deployed to provide a more densely instrumented sub-array in the center of
IceCube called DeepCore.  Each of these strings has 60 DOMs, 50 of which
are between depths of 2100 and 2450 m, which is below the main dust layer
at the South Pole.
The DeepCore sub-array is defined to include the bottom half of the central
string of IceCube as well as the lower DOMS on six surrounding standard strings.  
Typical spacing between strings in DeepCore is 75 m as compared to 125 m between
standard strings.  The spacing between DOMs on the 8 special strings is
7 m.  One of the main motivations is to provide
an inner fiducial volume surrounded by at least three rings of standard
IceCube strings and thirty layers of DOMs above to provide a veto against
atmospheric muons and allow for TeV neutrino astronomy in the Southern sky.
This and other goals of DeepCore are described in Ref.~\cite{DeepCore}.

\section{Flaring, transient and gamma-ray burst sources}

A correlation in space or time (or both) with a source observed
electromagnetically would enhance the likelihood that the observed
neutrinos are of astrophysical origin.  Two strategies are used in IceCube
to pursue this option.  One is to send alerts for follow-up by other
instruments when an apparently significant grouping of neutrinos is seen.
The potential signal could be two or more neutrinos from the same direction
within a short time window or a sequence of events from a targeted
source that exhibits flares in the electromagnetic spectrum, for example.  Currently
alerts are sent from IceCube to ROTSE and the Palomar Transient Factory
and SWIFT~\cite{Anna}.  Arrangements for sending alerts
to the Northern hemisphere gamma-ray telescopes MAGIC and
VERITAS are also in preparation.  These multi-messenger agreements lead to
the possibility that a neutrino signal could be associated with an
identified type of object, such as a supernova explosion or a
flaring AGN.  Limits obtained from the follow up with ROTSE~\cite{ROTSE}
are described in Ref.~\cite{Anna2}.

Going the other direction, IceCube can follow up potentially interesting
astrophysical events such as nearby supernovae~\cite{SN2008D} or flares from likely cosmic
accelerators~\cite{flaresearch}.  A general search
for short-term increases in rates of neutrinos from 
 flaring sources is also done~\cite{transient}.
The most important result achieved from a catalog of events is the search for 
neutrinos associated with gamma-ray bursts (GRB)~\cite{GRBprl}.  
Recently data sets from two years of IceCube while the detector
was still under construction (IC-40 and IC-59) have been combined to obtain
a significant limit on models~\cite{WB-GRB} in which GRBs are the main source of
extragalactic cosmic rays.  In total 215 GRBs reported
by the GRB Coordinated Network between April 5, 2008 and
May 31, 2010 in the Northern
sky were included in the search.  No neutrino was found during the
intervals of observed gamma-ray emission.

\begin{figure}[thb]
\begin{center}
\includegraphics[width=0.8\columnwidth]{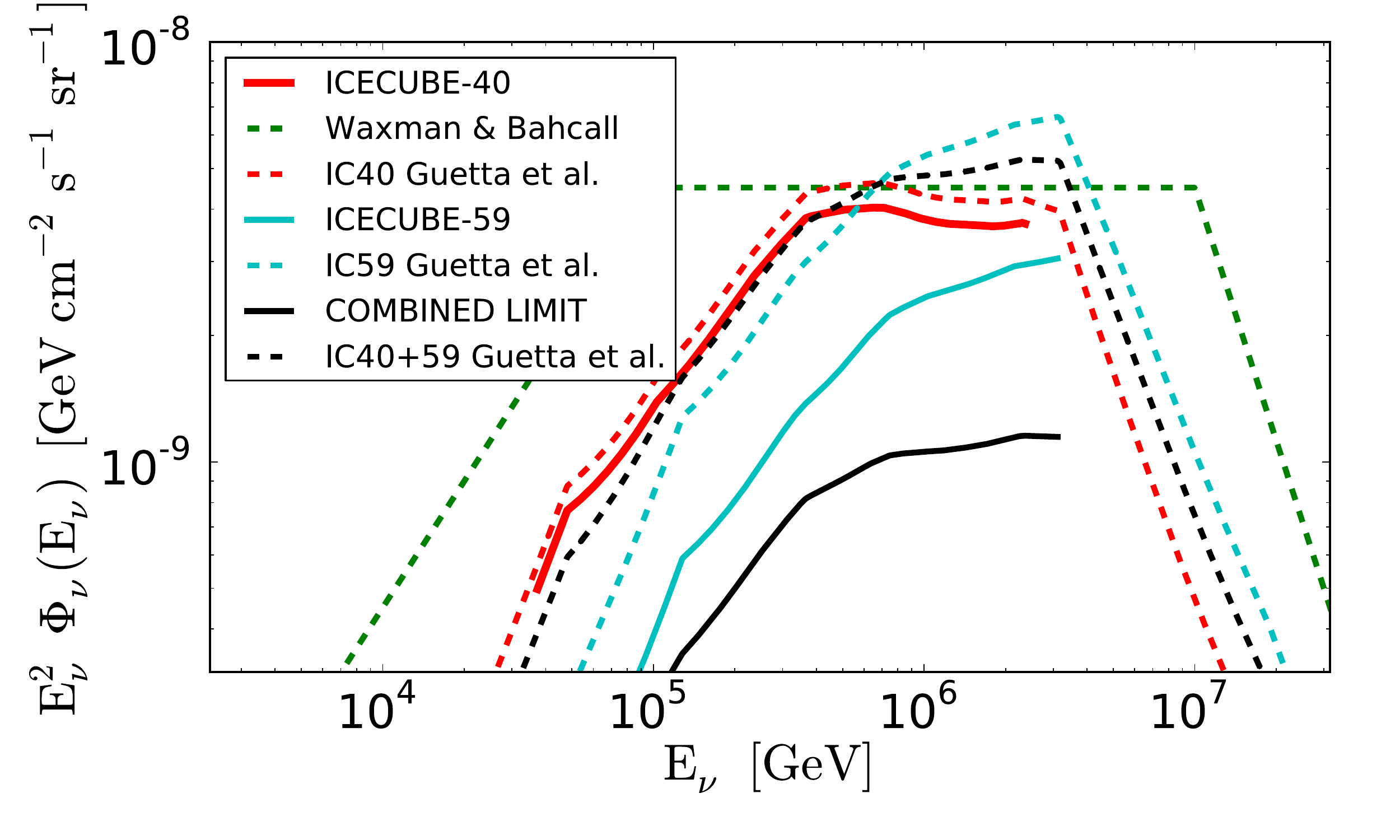}
\caption{Limits on neutrinos from GRB with expected
spectra calculated from Ref.~\cite{Guetta}.
Figure is from Ref.~\cite{Anna}.}
\label{GRB}
\end{center}
\end{figure}

To set limits on the model~\cite{WB-GRB}, the expected neutrino spectrum
was calculated for each burst based on parameters derived~\cite{Guetta}
from features in the spectrum of the GRB.  In particular,
a break in the observed photon spectrum marks the onset of photo-pion
production by accelerated protons
interacting with intense radiation fields in the GRB jet.  
The neutrinos come from the decay of
charged pions.  Given a predicted neutrino spectrum, the expected
number of events was calculated for each burst.  The normalization
of the calculation is provided by the assumptions that a fraction
of the accelerated protons escape and provide the ultra-high energy
cosmic rays.  In the simplest case, the UHECR are injected
as neutrons from the same photo-production processes in which the
neutrinos are produced~\cite{Halzen1}.  With this normalization, 8 neutrinos are
expected in 215 GRBs and none are observed.
Fig.~\ref{GRB} shows the resulting limits along with the
original prediction of Waxman and Bahcall~\cite{WB-GRB}.
A model-independent search was also carried out, simply looking
for neutrinos within $10^\circ$ in an expanding time window up to
several hours around the time of each burst.  Again, no correlated
neutrinos were found.

\section{All-sky survey with $\nu_\mu$}
\begin{figure*}[thb]
\begin{center}
\includegraphics[width=0.9\columnwidth]{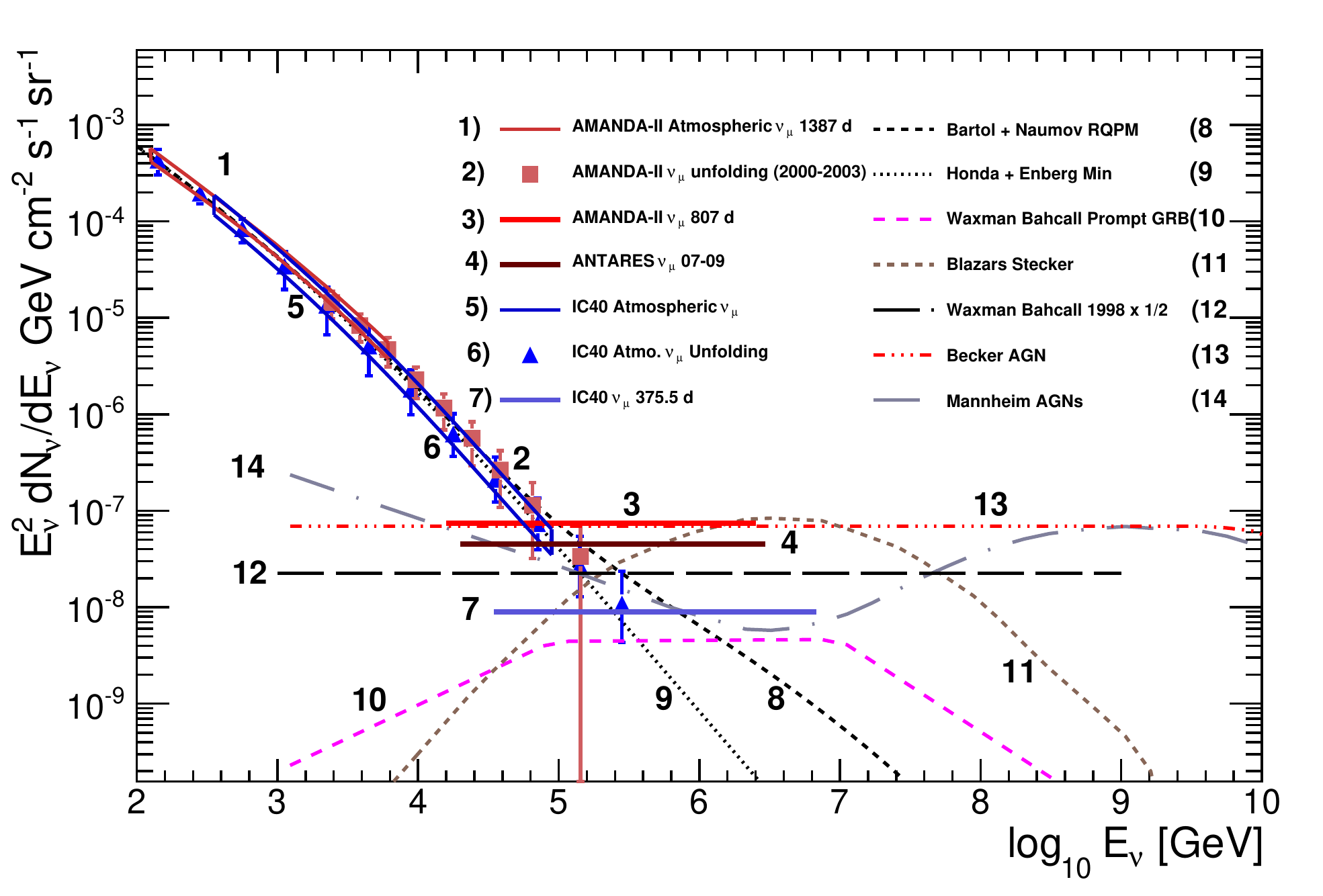}
\caption{Horizontal lines show limits on an $E^{-2}$ spectrum
of astrophysical muon neutrinos from AMANDA-II~\cite{AMANDA}, 
 Antares~\cite{Antares} and IceCube~\cite{IC40D}.  
The limits are shown along with measurements
of the flux of atmospheric muon neutrinos and anti-neutrinos.  The plot
is from Ref.~\cite{IC40D} where full references are given.
}
\label{fig2}
\end{center}
\end{figure*}

It is important also to search for an excess
of astrophysical neutrinos from the whole sky at high energy above 
the steeply falling background of atmospheric neutrinos.  The
Universe is transparent to neutrinos, so the flux of neutrinos
from sources up to the Hubble radius may be large~\cite{Lipari}.
A specific estimate based on the density of AGNs~\cite{Halzen}
is that the event rate from the whole sky should be
a factor 20 larger than from a single AGN.
Limits from AMANDA~\cite{AMANDA}, Antares~\cite{Antares}
and IceCube~\cite{IC40D} are shown in Fig.~\ref{fig2}.
The current limit from the 40-string version of IceCube 
is now below the original Waxman-Bahcall bound~\cite{WB}.

The limit is obtained by fitting the distribution of light
in the detector from upward moving muons to parameters
that describe an assumed neutrino spectrum consisting of
\begin{itemize}
\item ``conventional''atmospheric muon neutrinos from decay of charged pions and
kaons;
\item ``prompt'' atmospheric neutrinos, mainly from decay of charmed hadrons
with a spectral shape taken from Ref.~\cite{Enberg}; and
\item an assumed astrophysical flux with a hard $E^{-2}$ differential
energy spectrum.
\end{itemize}
The atmospheric neutrino flux used
in this analysis~\cite{IC40D} is a simple power-law extrapolation
of the calculation of Ref.~\cite{Honda06}.  
Its normalization is treated as a free parameter in fitting
the data in Ref.~\cite{IC40D}, which is shown as a slightly
curved band that extends from 0.33 to 84~TeV in Fig.~\ref{fig2}.
The other experimental
results on the high-energy flux of atmospheric $\nu_\mu+\bar{\nu}_\mu$
in Fig.~\ref{fig2} are from AMANDA~\cite{AMANDA1,AMANDA2} 
and IceCube-40~\cite{Warren}.  All the atmospheric neutrino spectra
shown here are averaged over angle.  The unfolding analysis of Ref.~\cite{Warren}
extends to $E_\nu\approx 400$~TeV.  
The integral limit on astrophysical neutrinos shown for IceCube-40 in Fig.~\ref{fig2}
assumes a hard, $E^{-2}$ spectrum.  For this reason,
the bound applies at much higher neutrino
energies (35 TeV to 7 PeV) than the observed spectrum of atmospheric neutrinos.

As discussed in Ref.~\cite{Gaisser-ricap}, one of the difficulties
of searching for an astrophysical contribution at the level of the
Waxman-Bahcall limit is that the atmospheric background in the
relevant energy range above $100$~TeV is not well known.  
Standard calculations of conventional atmospheric 
neutrinos~\cite{Bartol04,Honda06}
extend only to 10~TeV.  In addition, the level of prompt neutrinos,
which are expected to become important somewhere above $100$~TeV
is highly uncertain.  The current IceCube limit appears
already to rule out the highest prediction
for charm~\cite{RQPM}.  

\section{Implications}

The Waxman-Bahcall bound is an upper limit to the intensity of neutrinos which
holds if the neutrinos are produced in the same sources that produce
the extra-galactic cosmic rays.  For this to occur, the accelerated
protons must be able to escape from the sources, either directly or
as neutrons created by
$p\,+\,\gamma\rightarrow\Delta^+\rightarrow n \,+\,\pi^+$.  For GRB, escape (at least of
the neutrons) is likely because of the rapid expansion,
as discussed in Ref.~\cite{Halzen1}. 
In the case of AGNs,
Waxman and Bahcall~\cite{WB} show that the existence of TeV photons
from blazars guarantees that the density of electromagnetic
radiation is such that EeV nucleons can indeed escape, so the bound
applies to AGNs as well as GRBs.  If protons are trapped in
the acceleration region by the turbulent
magnetic fields needed to make the acceleration process work, 
then the bound may be related to an
estimate of the expected level of neutrino production. 
In this scenario, the high energy protons lose energy 
by photo-pion production, while the neutrons escape 
and decay to protons to
become ultra-high energy cosmic rays (UHECR). 
The neutrinos from the pion decay are then related by kinematics to the UHECR.  
Another consequence is that the UHECR would be protons.

This scenario could be realized in jets
of GRB and of AGN if acceleration occurs in
internal shocks in the jets.  As IceCube limits become
increasingly strong, this class of models is constrained.
A generic alternative could be that the UHECR are
accelerated outside the jets, for example at the
termination shocks of AGNs.  In this case the
composition of the extragalactic cosmic
rays would depend on the ambient medium, and the level of
neutrino production would be contingent on the 
density of the surrounding medium and correspondingly low.

\begin{figure*}[thb]
\begin{center}
\includegraphics[width=0.9\columnwidth]{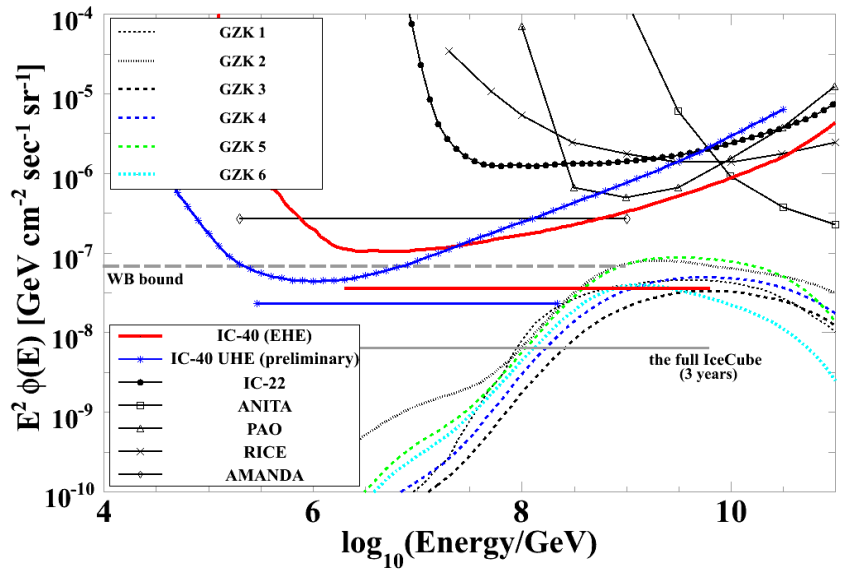}
\caption{Collection of limits on cosmogenic and ultra-high
energy neutrinos.  The plot
is based on Ref.~\cite{EHE} where full references are given.
The extra curve included here, labeled {\em IC-40 UHE (preliminary)}
is from Ref.~\cite{Henrik}.
}
\label{UHEnu}
\end{center}
\end{figure*}

\section{Cosmogenic neutrinos}
Independent of the level of neutrino production in cosmic sources,
there will be production of neutrinos from
$$p + \gamma_{\rm CMB} \rightarrow\Delta^+\rightarrow n+\pi^+\rightarrow \nu_\mu
\;(\rightarrow p+\pi^0\rightarrow \gamma\,\gamma)$$
provided only that there are protons above the threshold 
of $\sim5\times 10^{19}~eV$ from sources distributed
throughout the Universe.  Photons are produce in the
neutral pion channel of the same process.  These photons would
undergo cascading and contribute to the background
of diffuse gamma-rays measured by the Fermi Satellite~\cite{Fermi}, 
which leads to an upper limit on
the spectrum of cosmogenic neutrinos.  This limit
is now at the level where no more than approximately one event
per year would be expected in the full IceCube~\cite{Ahlers3}.

Limits from IC-40~\cite{EHE} and other detectors are collected
in Fig.~\ref{UHEnu}.  The limits are for the sum of 
three flavors assuming equal contributions from each flavor.
Both differential and integral limits
are shown.  Differential limits are obtained by assuming an
$E^{-2}$ spectrum over a logarithmic bin of energy 
and calculating the limit for the each bin.
The shapes of the curves then indicate the energy
response of each detector.
The limits for a full $E^{-2}$ neutrino spectrum are shown
as straight lines in the plot.  The most precise result
is to calculate the expected signal for each flux model
and state the corresponding limit or expectation.  For example,
for the flux of Ref.~\cite{Ahlers3} the expected number
of events is 0.43 for IC-40 detector, which was half the
size of the completed IceCube.

The ANITA detector~\cite{ANITA} is designed to look for
neutrino interactions in the ice while flying in a circumpolar
pattern in long duration balloon flights above Antarctica.
From its altitude of $\approx 35$~km, an array of antennas
scans $\approx10^6$~square kilometers of ice looking for
radio pulses produced by
Cherenkov radiation from the cascade generated by
the neutrino interaction in the ice.  The radio pulse from
a downward slanting neutrino would be
reflected from the ice-rock interface and therefore identifiable
by its vertical polarization.  One neutrino candidate was
detected in the second flight of ANITA~\cite{ANITA-erratum},
which was aloft for 31 days.
ANITA also sees horizontally polarized events which are
interpreted as air showers with cosmic-ray
energies of $\sim10^{19}$~eV that generate a radio
signal by geo-synchrotron radiation~\cite{ANITA-CR}.  A third flight of ANITA is
planned for 2012/13.

The Auger detector is sensitive to neutrinos with energies
in the EeV range~\cite{Auger}.  The signal of a neutrino would be a
horizontal air shower with a normal electromagnetic component
from a neutrino interaction near the detector.
Horizontal air showers produced by cosmic rays would have
originated far away so that the electromagnetic component
would be absorbed leaving mostly muons at the detector.
With a surface area of 3000 km$^2$ the volume of air
directly over the array is equivalent to $\approx30$~km$^3$ of water.
However, the array is most sensitive (by a factor of 
about five~\cite{AugerICRC})
to Earth skimming charged current interactions of $\nu_\tau$ 
in which a tau emerges from the nearby mountain and initiates 
an air shower over the array when it decays.
The expected number of cosmogenic neutrino events expected
in several year is 0.71 for the Earth skimming channel
and 0.14 for the near horizontal air shower channel~\cite{AugerICRC}.

\section{Cosmic-ray physics with IceTop}

\begin{figure}[thb]
\begin{center}
\includegraphics[width=0.8\columnwidth]{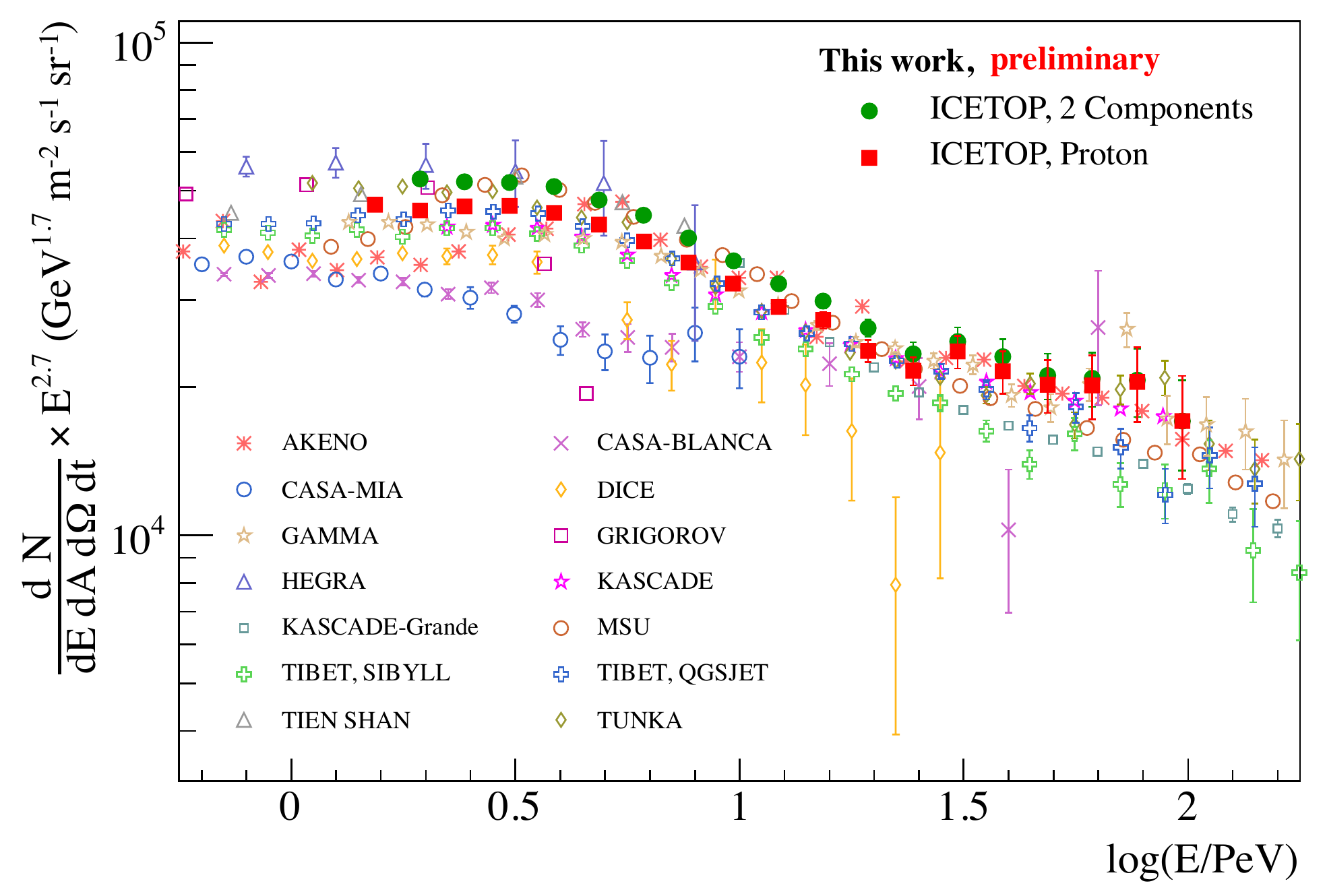}
\caption{Preliminary measurement of the primary
energy spectrum with IceTop in 2007 with 26 stations deployed.
Figure from Ref.~\cite{FabianThesis}.}
\label{Espectrum}
\end{center}
\end{figure}

The surface array of IceCube consists of 81 stations with two
ice Cherenkov tanks separated from each other by 10 meters.
Each tank contains two downward facing DOMs with the
PMT half of the DOM frozen into the ice.  During deployment,
the tanks were filled with water and then frozen in a controlled
process to achieve clear ice free of bubbles.  Tanks are
lined with a diffusely reflecting material which scatters
the Cherenkov light from charged particles back up toward
the ice surface and the PMTs.  The typical signal from
a muon passing through the tank is 130 photo-electrons.
One PMT is set at a lower gain than its partner so that
the dynamic range of a tank is $\approx 0.1$ to $1000$ vertical
equivalent muons.  Figure~\ref{Espectrum} is a preliminary
measurement of the spectrum with IceTop in 2007 when there
were 26 stations~\cite{Kislat}.

The area enclosed by IceTop is approximately one km$^2$, and the
mean atmospheric overburden at the South Pole is $680$~g/cm$^2$.
Given its total area and the average spacing between stations of $125$~m,
IceTop covers an energy range from $300$~TeV to just
above $1$~EeV.  The aperture of the 3-dimensional array
formed by and IceTop and the deep part of IceCube is $\approx{1\over 3}$km$^2$sr.
Together, the two parts of IceCube form a three-dimensional
air shower array.

\begin{figure}[b!]
\begin{center}
\includegraphics[width=0.4\columnwidth]{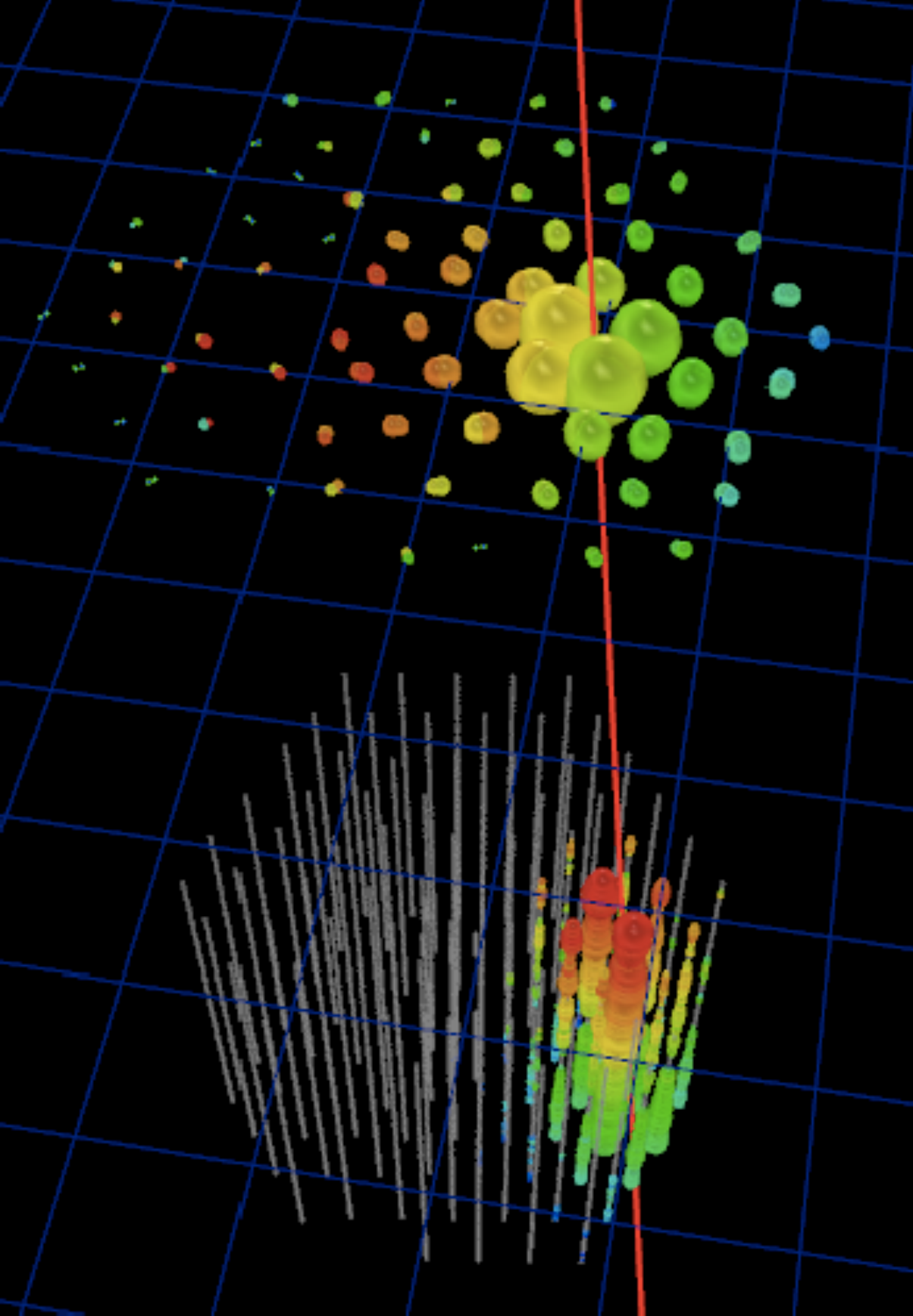}
\caption{Example of a coincident event contained
in both parts of IceCube.}
\label{event}
\end{center}
\end{figure}

Figure~\ref{event} shows a coincident event that is 
contained in both parts of the IceCube.  The event
has an energy of $\approx30$~PeV as estimated from the pulses in IceTop.
Such at event would produce 35 to 100 muons with sufficient
energy at production in the atmosphere to penetrate through
IceCube.  The actual number fluctuates from event to event,
but these are average values expected for protons or iron nuclei
respectively.  A major goal of cosmic-ray studies with
IceCube is to use the ratio of signal in the deep detector
to shower energy estimated by IceTop to measure the
primary composition up to the EeV range.  A preliminary 
analysis with 21 days of data taken with 
IC-40~\cite{Kath} indicates an increase in the mean
mass of the primary cosmic rays between 0.3 and 30 PeV.
 Analysis of a large sample of data from the completed detector
to exploit its full capability depends on correcting for
systematic seasonal effects, which is underway.

\section{Conclusion and Outlook}

So far the only extraterrestrial neutrinos detected are
from Supernova 1987A~\cite{Kam,IMB} and from the Sun~\cite{SNO,SuperKsolar}.
Early predictions for high levels of astrophysical neutrinos
at high energy have not been realized.  Current limits from
IceCube are beginning to exclude some models in which production
of neutrinos is intimately connected with acceleration of the
majority of ultra-high energy cosmic rays from extra-galactic sources.
At the same time, the sensitivity of IceCube is approaching the level
where signals are expected from some galactic sources after several
years~\cite{Halzen}.  Realization of the proposals for KM3NeT~\cite{KM3NeT}
and (or) GVD~\cite{GVD} in the Northern hemisphere 
would provide full-sky coverage with cubic kilometer sensitivity
for astrophysical neutrinos in the TeV range and above. 
In the meantime, ANTARES and Baikal continue to provide some
level of complementary coverage as indicated in Fig.~\ref{skymaps}.

An important aspect of IceCube from the point of view of
high energy physics is the possibility to use the atmospheric
neutrino beam to search for hints of new particle physics~\cite{Helbing}
that might show up as anomalies in the angular dependence 
or the energy spectrum of atmospheric neutrinos.  This is possible
because the large size of IceCube gives an unprecedented
rate of some 100,000 neutrinos per year in the TeV range and 
above.  Limits on violation of Lorentz invariance~\cite{Warren1}
and on certain models of sterile neutrinos~\cite{Razzaque}
have already been published.  The energy dependence of the
neutrino cross section above accelerator energies is also
of interest in this connection.

Another consequence of the large size of IceCube is the huge
sample of cosmic-ray induced atmospheric muons.  The event
rate of almost 100 billion $\sim$TeV muons per year has already
led to measurement in the Southern sky of cosmic-ray anisotropy
at the level of $10^{-4}$~\cite{anisotropy1} and to new measurements
of its small-scale structure~\cite{anisotropy2} and energy dependence~\cite{anisotropy3}.
Statistics are high enough to measure the spectrum of atmospheric muons
to $100$~TeV and somewhat higher~\cite{Patrick}.  This opens
the possibility of measuring the prompt neutrinos from charm
by a combination of angular dependence and seasonal variations~\cite{DG}.

Other capabilities of IceCube include detection of neutrinos from galactic
supernova explosions and sensitivity to neutrinos from dark matter annihilation.
Rates above threshold in the DOMs are monitored continuously.  A galactic supernova
would show up as a sharp increase in the counting rate due to light
produced near the DOMs by many interactions of $\sim$10~MeV neutrinos~\cite{Lutz,SNpaper}.
This is possible because of the low counting rate ($\approx 500$~Hz) of DOMs in the ice.
Similar monitoring of scalar rates in IceTop DOMs has already led to
the observation of particles at ground level associated with a solar flare~\cite{Solar}.
A sufficiently strong GRB could also show up as an increase in IceTop single tank rates.

Indirect searches for dark matter are done in two ways in IceCube.  Neutrinos from WIMP annihilation in
the Sun provide the most stringent limits on the spin-dependent WIMP-proton
scattering cross section~\cite{SolarWIMP}.
It is also possible to look for neutrinos from WIMP annihilation the Galactic halo (currently
by looking for an angular excess in the edge of the galactic center region that
is visible in the Northern sky~\cite{HaloWIMP}).

An important aspect of IceCube is its ability to detect all flavors of neutrinos~\cite{Joanna}.
Although the largest event rate at high energy will be from charged current interactions
of $\nu_\mu$ because the long muon range, the other flavors are particularly 
important for several reasons.  The background of atmospheric $\nu_e$
is significantly lower at high energy than $\nu_\mu$.  To be identified,
$\nu_e$ must interact inside the detector, which means their energies
will be measured more accurately.  Production of $\nu_\tau$
by cosmic-ray interactions in the atmosphere is negligible, but, because of
oscillations, they will be at the same level as $\nu_\mu$ and $\nu_e$ from
astrophysical sources.  Identification of events as $\nu_\tau$ would therefore
be a strong indication of astrophysical origin.

Flavor identification hinges on identifying cascades in the detector, where
a cascade is defined as a nearly isotropic burst of light.  High energy muons
produce cascades by bremsstrahlung and other stochastic loss processes, but they
can be identified by the associated track.  Charged current interactions
of $\nu_e$ as well as neutral current interactions of all flavors make isolated
cascades.  Cascades at the rate expected from atmospheric neutrinos 
have already been detected in the sub-TeV range with DeepCore~\cite{Ha}.
Depending on their energy, $\nu_\tau$ would be identified as a single
(perhaps elongated) cascade, a ``double bang" event~\cite{Pakvasa} or a cascade
associated with the track of a $\tau$-lepton either before or after the cascade
depending on whether the cascade is from the lepton decay or the neutrino interaction.
The $\tau$ track would have lower brightness than a muon track of similar energy.
In addition, the $\nu_\tau$ channel is amplified by regeneration
in the Earth through $\nu_\tau\rightarrow\tau\rightarrow\nu_\tau$~\cite{Halzen2,Beacom}.
It is interesting to note that the cascade-like events generated primarily by $\nu_e$ and 
$\nu_\tau$ give significant contributions to the limits obtained for $E_\nu > 1$~PeV
in IceCube~\cite{EHE,Henrik}, as shown in Fig.~\ref{UHEnu}.

For the future there are new efforts that push toward larger exposure at the highest
energies as well as plans for increasing the sensitivity at low energy.
The KM3NeT project aims for a substantial increase in sensitivity compared to IceCube, with
emphasis on the galactic center region.  The Askaryan Radio Array (ARA) project has the aim of
building a detector capable of detecting about 100 cosmogenic neutrinos
at the level allowed by the diffuse gamma-ray limit of Fermi~\cite{Ahlers3}.
The first test set up for ARA was deployed next to IceCube in January, 2011.
Radio antennas will be deployed in shallow holes ($\approx 200$~m) starting in the 
current (2011/2012) Antarctic seasons to detect Askaryan radio emission from neutrinos~\cite{ARA}.
There is also a proposal for a radio air shower test array (RASTA) to determine
the feasibility of expanding the surface air shower coverage by an order of magnitude~\cite{RASTA}.

In the low-energy direction, there is a proposal to lower the energy
threshold of the current DeepCore to $<10$~GeV by deploying 18 to 20 strings
inside the DeepCore volume~\cite{Ty}.  This extension (called PINGU) would enhance the capability
of IceCube for neutrino oscillation studies and for low energy neutrinos
from the galactic center region in the Southern sky.  There is also ongoing
discussion of the possibility of making megaton scale proton decay detector
in the deep center of IceCube, which would require much denser coverage~\cite{Darren}.

\vspace{-.5cm}

\acknowledgments
\vspace{-.2cm}
This work is support in part by the U.S. National Science Foundation.  I thank Gisela Anton
and Paschal Coyle for providing information about Antares.  I am grateful to several
colleagues in IceCube for their contributions of figures and for improvements in the text.

\bibliography{references}

\end{document}